\documentstyle[epsfig]{aipproc}

\begin{document}
\title{Infrared Galaxies} 

\author{Andrew W. Blain}
\address{Cavendish Laboratory, Madingley Road, Cambridge, CB3 0HE, UK} 

%\lefthead{LEFT head}
%\righthead{RIGHT head}
\maketitle

\begin{abstract}
To study large-scale structure in the Universe a full census of the contents 
are required. This is even more important when the processes of galaxy formation 
are being investigated. In the last year the population of distant 
galaxies that emit most of their energy in the infrared waveband have been 
studied in unprecedented detail. The intensity of background radiation 
at wavelengths between 3\,mm and 10\,\micron\ has been determined to 
within a factor of about 2, and much of this background radiation has been 
resolved into individual galaxies. These galaxies are largely hidden 
from view in the optical waveband by absorption due to interstellar dust. By 
combining this new knowledge with the ever growing body of information 
gathered using optical telescopes, the process of galaxy formation is slowly 
being revealed. Great opportunities for studying the distant Universe are
promised by advances in instrumentation in the millimeter (mm), submm and 
far-infrared wavebands over the next decade. 
\end{abstract}

\begin{figure}[ht]
\centerline{\psfig{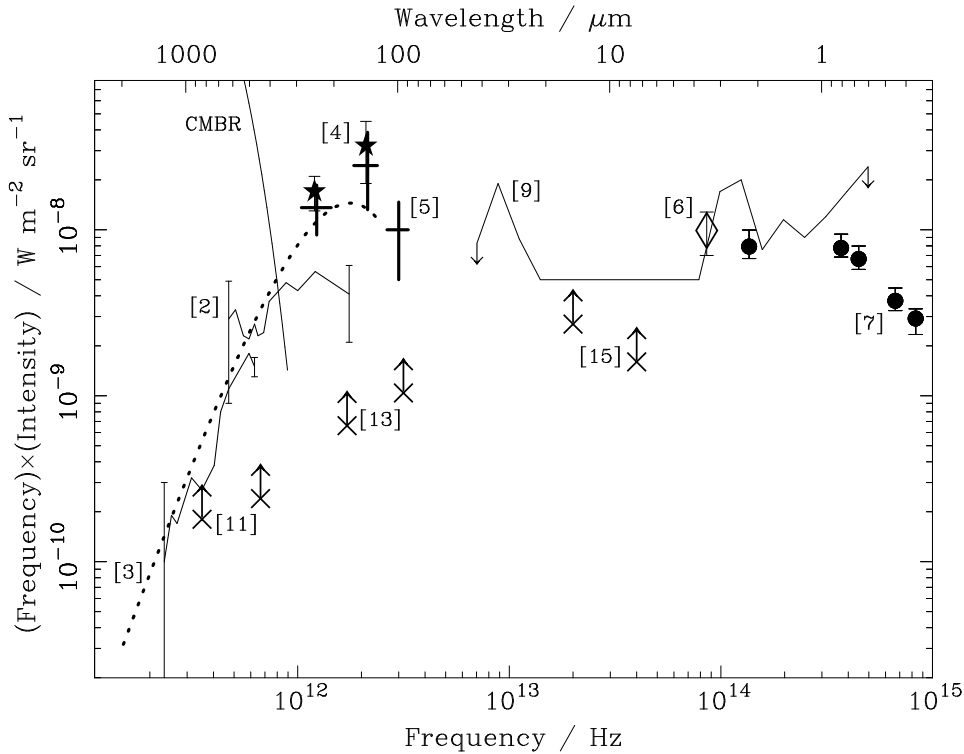}}
\vspace{10pt}
\caption{ {\baselineskip = -150pt The background radiation intensity from 
the mm to optical waveband. 
References are numbered. Only [2] was obtained before 1998. 
Note that at wavelengths of 15 and 850\,\micron\ [11,15]
most of the background has probably been resolved into its 
constituent galaxies.} } 
\label{fig1} 
\end{figure} 

\section*{Introduction}

Progress in observational cosmology has been unprecedentedly rapid over 
the last five years, with the advent of the {\it Hubble Deep Fields}, 
Lyman-break galaxies and the Keck telescopes. However, there is 
still no detailed understanding of how and when galaxies formed. In the optical 
waveband progress should continue, due to the large number of 8-m-class 
telescopes being commissioned, the increasing sophistication of multi-object 
spectrographs and new wide-field digital surveys.

A large amount of the energy output of galaxies is emitted in the far-infrared 
waveband, as revealed by the {\it IRAS} satellite \cite{SM}. This is thermal 
emission from interstellar dust grains, which efficiently absorb blue and 
ultraviolet photons, whether from young high-mass stars or active galactic 
nuclei (AGN). Heated to temperatures of a few tens of Kelvin, the rest-frame 
emission spectrum of the dust peaks at wavelengths close to 100\,\micron. 
In the local Universe, a comparable amount of energy is emitted in the optical 
and far-infrared wavebands. However, because the {\it IRAS} survey extended 
only to redshifts $z \simeq 0.2$, until recently it was impossible to investigate 
the reprocessed dust radiation from distant galaxies directly, either individually 
or as an integrated background \cite{puget}. 

\section*{Backgrounds and discrete sources} 

Within the last year our knowledge of the quantity of energy re-radiated 
by dust 
in distant galaxies has increased dramatically (see Figure 2). Using the FIRAS 
and DIRBE instruments on the {\it COBE} satellite, and the {\it HST (Hubble 
Space Telescope)}, the intensity of extragalactic background radiation has been 
determined to within a factor $\sim 2$ between wavelengths of 1\,\micron\ 
and 2\,mm \cite{fixsen,sfd,hauser,DA,pozzetti,bernstein}. Upper limits to the 
background intensity from $\gamma$--$\gamma$/e$^+$e$^-$ pair production 
along the line of sight to nearby AGN have also been tightened\cite{SF}. Two 
years ago, there was only a single background estimate \cite{puget}. The new 
observations represent a tremendous advance. 

The first results of direct surveys to detect the faint galaxies that contribute to 
the infrared background radiation intensity have also been published this year. 
Varying  fractions of the background have been resolved into individual galaxies 
using three sets of instruments. At long wavelengths of 850 and 450\,\micron, 
the SCUBA bolometer array camera\cite{holland} at the JCMT (James Clerk 
Maxwell Telescope) has been used to image the sky behind distant clusters of 
galaxies. By exploiting their gravitational lensing effect, most of the 
850-\micron\ background has been resolved into individual galaxies 
\cite{SIB,BKIS}. At shorter wavelengths of 175 and 95\,\micron, that are 
inaccessible from the ground, the {\it ISO} satellite has resolved about 
10\% of the background \cite{kawara,pugetiso}. Extremely faint 
galaxies at mid-infrared wavelengths of 15 and 7\,\micron\ have also
been determined using {\it ISO} by exploiting gravitational lensing \cite{altieri}. 
The lower limit to the background derived from the 15-\micron\ observations is 
close to the $\gamma$--$\gamma$ upper limit \cite{SF}, suggesting 
that a large fraction of the 15-\micron\ background has been resolved. Several 
other submm-wave SCUBA surveys have been carried out 
\cite{barger,hughes,eales}, and an interesting upper limit to the surface density 
of galaxies has been imposed at a wavelength of 2.8\,mm \cite{wilner}. 

The prime motivation for submm-wave galaxy surveys is their ability to select 
high-redshift galaxies preferentially. The form of the dust emission spectrum of 
galaxies is broadly similar to the envelope of the background limits shown in 
Figure~1. If a high-redshift galaxy is observed at a wavelength of several 
hundred microns, then its rest-frame spectrum is sampled at a wavelength 
considerably closer to the peak of the emission spectrum. As a result, the 
detected flux density is increased; this effect is described as a negative 
$k$-correction. In some cases this $k$-correction is sufficiently large to 
overcome the inverse square law dimming of the galaxy with increasing redshift, 
and so the flux density is almost constant over reasonable redshifts 
$0.5 \lesssim z \lesssim 10$. This effect is illustrated in Figure~2, for a galaxy 
with an intrinsic luminosity of $5 \times 10^{12}$\,L$_\odot$ and a dust 
spectrum that provides a good fit to the properties of the background and 
surface density of dusty galaxies \cite{BSIK}. Note that the $k$-correction is 
important only at observing wavelengths longer than about 200\,\micron. 

\begin{figure}[ht]
\centerline{\psfig{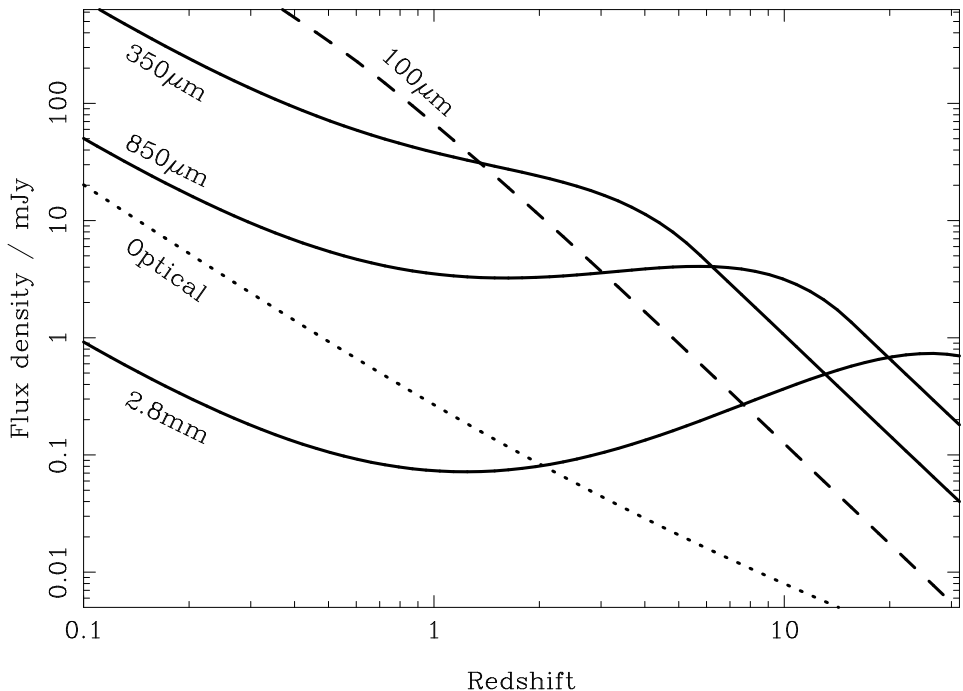}}
\vspace{10pt}
\caption{
The flux density expected from a galaxy with a fixed rest-frame 
spectrum as a function of redshift. In the optical (or radio) and mid-infrared 
wavebands, the received energy falls steadily with redshift. However, at 
wavelengths longer than 200\,\micron, the effect of redshifting the steep 
far-infrared dust spectrum counteracts the dimming effect of the inverse square 
law, and can lead to a brightening of the 
template galaxy as the redshift increases. This effect is responsible for the 
unique sensitivity of mm/submm-wave surveys to high-redshift galaxies/AGN.} 
\label{fig2} 
\end{figure}

The spectrum of a dusty galaxy is a power-law at wavelengths longer than a few 
hundred microns with a temperature dependent turn-over at shorter 
wavelengths. Measurements of flux density both above and below the
wavelength of the turn-over are thus required to determine the dust 
temperature, which in turn fixes the total luminosity. Hence, there is a very 
strong case for carrying out galaxy surveys at long submm wavelengths, to 
exploit the $k$-correction effect, and then conducting follow-up observations 
of the detected galaxies at shorter wavelengths. For a galaxy 
discovered in a survey at a wavelength of 500\,\micron, a sensitive follow-up 
observation at 100\,\micron\ is much more useful than one at 1000\,\micron. 

The recent detection of well defined populations of distant dusty galaxies, 
especially at wavelengths of 850 and 15\,\micron\ \cite{SIB,altieri}, confirms 
that a population of dusty galaxies exists at large redshifts. The background 
radiation intensity is dominated by galaxies with redshifts $z \simeq 1$, and so 
the measured background spectrum is not very useful for probing the detailed 
properties of more distant galaxies. In contrast, the properties of the galaxies 
detected by SCUBA, especially their redshift distribution \cite{BSIK}, will be 
extremely helpful in this respect. 

The best studied galaxy discovered by SCUBA, SMM\,J02399$-$0136, 
is at $z=2.8$ \cite{ivison,frayer}. Most of the 
other galaxies detected by SCUBA have plausible identifications on deep 
{\it HST} images \cite{SIBK}, and so are likely to lie at $z \lesssim 5$. Thanks 
to the power of the 10-m Keck telescopes, attempts to determine redshifts 
for 16 SCUBA-selected galaxies \cite{SIBK} are making 
considerable progress \cite{bargercowie}, despite the uncertain fraction of 
their luminosity that is absorbed by interstellar dust. 

\begin{figure}[ht]
\centerline{\psfig{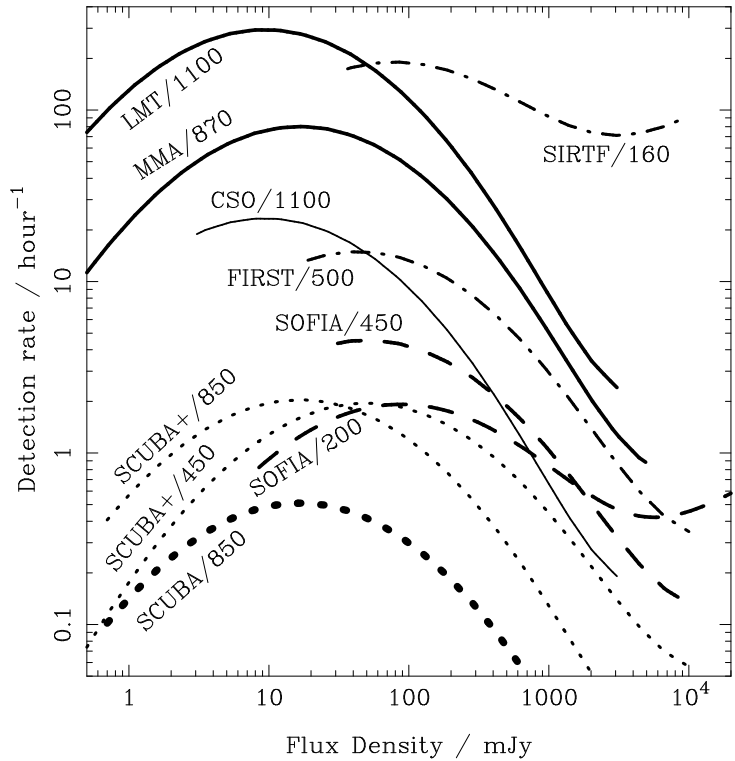}}
\vspace{10pt}
\caption{The detection rate expected in surveys using future 
mm/submm/far-infrared instruments. The lines are labeled with the instrument 
name and the survey wavelength in microns. The left and right end of the lines 
correspond to the flux density of the confusion limit [25] 
and the brightest 
source expected over the whole sky respectively. Note that different points 
on the same line correspond to independent surveys with different flux density 
limits and scanning rates. The detection rate expected for {\it SIRTF} at 
a wavelength of 70\,\micron\ 
is even greater than that at 160\,\micron, although the 
typical redshift of the detected galaxies will be less. 
A $5\times 10^{12}$-L$_\odot$ galaxy at $z=3$ is expected to produce flux
densities of 0.7, 1.5, 6.2, 8.9 and 6.1\,mJy at wavelengths of 1100, 850, 
450, 200 and 160\,\micron\ respectively.}
\label{fig3}
\end{figure}

\section*{Future instrumentation} 

{\it COBE} has done an excellent job of mapping the extragalactic infrared 
background. However, there is still a great deal of progress to be made in the 
detection and study of a large sample of galaxies at wavelengths between 
several mm and 5\,\micron. The value of the recent contribution of SCUBA
\cite{holland} cannot be overstated; however, 
SCUBA is just the first submm-wave instrument with the sensitivity required to 
map the distant Universe. SCUBA currently detects a galaxy every few hours, 
and at 850\,\micron\ is limited by source confusion at its 14-arcsec resolution 
in a few tens of hours of integration in a single field\cite{BIS}. 
Great progress is being made in 
mm/submm-wave instrumentation, improving both mapping speed and angular 
resolution. For example, the BOLOCAM instrument \cite{glenn}, destined for the 
10-m Caltech Submillimeter Observatory (CSO) and the 50-m Large Millimeter 
Telescope (LMT), should increase the detection rate by a factor of 100--1000 
(see Figure 3). Large ground-based interferometer arrays, such as the MMA, 
will obtain sub-0.1-arcsec resolution submm-wave images of galaxies, and 
will be able to detect unknown galaxies at a rate similar to that of the LMT. 
This resolution will be crucial for following up all observations made using 
other telescopes at coarser angular resolution. To observe at wavelengths 
shorter than several hundred microns, air- or space-borne instruments are 
required, and are being planned and constructed. 

%The HAWC bolometer camera on the 2.5-m SOFIA airborne observatory 
%should be available within 4 years, and in 9 years the {\it Planck Surveyor} 
%satellite will provide an all-sky map containing at least as rich a variety of new 
%objects as that obtained by {\it IRAS} 15 years ago. The 3.5-m {\it FIRST} 
%satellite should be available at about the same time as {\it Planck Surveyor}. 
%At wavelengths shorter than 200\,\micron, {\it SIRTF} will be in 
%service on the same timescale as SOFIA, and will make detailed observations 
%of the galaxies and quasars that will hopefully be detected by the 
%{\it WIRE} satellite as this volume goes to print. 

The best strategy for exploiting the great improvements in survey speed and 
angular resolution of the full range of future mm/submm/far-infrared instruments 
cannot yet be finalized, and will remain subject to modification until they 
all are in service with verified sensitivities. 
Several points about the future of submm-wave galaxy 
surveys are clear, however: 
\begin{enumerate}
\item The space-borne 0.3-m {\it WIRE} and 0.85-m {\it SIRTF} will provide 
extremely large samples of galaxies/AGN at far-/mid-infrared 
wavelengths. The areas of their surveys would be ideal 
targets for future surveys at longer wavelengths. 
\item The 2.5-m SOFIA airborne observatory will be very timely, and sufficiently 
flexible and adaptable in operation, to provide the initial far-/mid-infrared 
follow-up observations of galaxies detected in the mm/submm waveband 
using SCUBA and BOLOCAM. 
\item Within 10 years, submm-wave surveys using the LMT, MMA, 
{\it Planck Surveyor} and 3.5-m {\it FIRST} will provide huge samples of 
submm selected high-redshift galaxies/AGN. The {\it Planck} all-sky map
will contain at least as rich a variety of new objects as that obtained by 
{\it IRAS} 15 years ago. 
\item All detailed follow-up observations, and any morphological studies, will 
require the sub-arcsecond angular resolution of a large interferometer array.
\end{enumerate} 
Once a large sample of 10$^5$ or more galaxies/AGN are obtained using these 
instruments, the well developed techniques of optical observational 
cosmology, such as the analysis of distribution functions, can be applied to 
investigate the evolution of galaxies and large-scale structure, 
independent of the effects of dust obscuration. 
%subject to 
%selection effects independent of those in the optical waveband. 
\section*{Gravitational lenses} 

There is another positive feature of galaxy surveys in the mm/submm waveband.
Because the chance of a foreground galaxy being directly aligned on the 
line of sight to a galaxy -- and thus a strong magnification being produced -- 
increases with redshift, and there is a unique high-redshift selection bias 
in submm-wave surveys, the probability of 
submm detected galaxies being gravitationally lensed is greater as 
compared with other wavebands \cite{B96}. The most relevant instrument 
for a lens survey is {\it Planck Surveyor}, which will detect the brightest 
lenses over
the whole sky \cite{lens}. A large sample of gravitational lenses offers the 
opportunity both to study galaxies that would otherwise be too faint to detect, 
and to investigate the geometry of the Universe. In order to select the 
subsample of the detected galaxies that display the distorted and multiply 
imaged structures typical of a lensed galaxy on arcsecond angular scales, 
MMA imaging will again be 
essential. 

\section*{Conclusions} 

Within the last year, the intensity of extragalactic background radiation from the 
mm to the near-infrared waveband has been determined for the first time. 
This energy was released at shorter wavelengths by star-forming galaxies and 
AGN during the process of galaxy formation, and has been reprocessed by 
interstellar dust. The coincident detection of populations of distant dusty 
galaxies/AGN at six wavelengths from 850 to 7\,\micron\ has provided much 
more information about the properties of distant infrared galaxies. The 
population of galaxies discovered using SCUBA at 850\,\micron\ is particularly 
interesting. Most of the 850-\micron\ background radiation intensity 
detected by {\it COBE} 
can be accounted for by the existing SCUBA galaxies, indicating 
that dust-enshrouded high-redshift galaxies/AGN were rare, but very luminous. 
The next generation of mm/submm-wave instruments -- ground-based, air- 
and space-borne -- will provide much larger samples of distant infrared 
galaxies. The array of data they provide will be comparable in both volume and 
quality with that obtained in the optical/near-infrared waveband. 

\section*{Acknowledgments} 

This work has benefited greatly from observations using SCUBA at the 
JCMT in collaboration with Ian Smail, Rob Ivison and Jean-Paul Kneib. 
I thank Jackie Davidson, Jason Glenn, Malcolm Longair, 
Priya Natarajan et les garcons SCUBA.

\end{document}